\input phyzzx

\def\IR{{\hbox{{\rm I}\kern-.2em\hbox{\rm R}}}}
\def\IB{{\hbox{{\rm I}\kern-.2em\hbox{\rm B}}}}
\def\IN{{\hbox{{\rm I}\kern-.2em\hbox{\rm N}}}}
\def\IC{{\ \hbox{{\rm I}\kern-.6em\hbox{\bf C}}}}

\def\IZ{{\hbox{{\rm Z}\kern-.4em\hbox{\rm Z}}}}
\def\to{\rightarrow}

\def\underarrow#1{\vbox{\ialign{##\crcr$\hfil\displaystyle
{#1}\hfil$\crcr\noalign{\kern1pt
\nointerlineskip}$\longrightarrow$\crcr}}}
% use of underarrow
%A~~~\underarrow{a}~~~B
%

\def\ltorder{\mathrel{\raise.3ex\hbox{$<$}\mkern-14mu
             \lower0.6ex\hbox{$\sim$}}}
\def\lesssim{\mathrel{\raise.3ex\hbox{$<$}\mkern-14mu
             \lower0.6ex\hbox{$\sim$}}}

%The usage is $X \catquot G$.

\font\Lina=cmbx10 scaled \magstep2
\font\Hugo=cmbx10 scaled \magstep3

\input phyzzx
\overfullrule=0pt
\tolerance=5000
\overfullrule=0pt
\twelvepoint

\twelvepoint

%\rightline{CINVESTAV-FIS 18/95}
%\rightline{January, 1996}
\titlepage
\vskip 2truecm

\centerline{\Hugo A Hopf algebra structure in self-dual gravity}
\vglue-.5in
\author{H. Garc\'{\i}a-Compe\'an
\foot{E-mail: compean@fis.cinvestav.mx}, L.E. Morales \foot{ E-mail: lem@fis.cinvestav.mx} and J.F. Pleba\'nski \foot{E-mail: pleban@fis.cinvestav.mx}}
\medskip
\address{Departamento de F\'{\i}sica
\break  Centro de Investigaci\'on y de 
Estudios Avanzados del IPN.
\break Apdo. Postal 14-740, 07000, M\'exico D.F., M\'exico.}
\bigskip
\abstract{The two-dimensional non-linear sigma model approach to Self-dual 
Yang-Mills theory and to Self-dual gravity given by Q-Han Park is an 
example of the deep interplay between two and four dimensional physics. 
In particular, Husain's  two-dimensional  chiral model approach to 
Self-dual gravity is studied. We show that the 
infinite hierarchy of conservation laws associated to the Husain model 
carries implicitly a hidden infinite Hopf algebra structure.}


%\noindent
%{\it Key words}: Self-dual Gravity, Symplectic Diffeomorphism Lie 
%Algebra, Hopf Algebra.
  
%\noindent
%1991 MSC: 81 V 17, 17 B 65, 17 B 66, 16 W 30

\endpage

\chapter{\bf Introduction}

Self-dual gravity (SdG) is a very interesting arena to understand the 
interplay between the physics and/or mathematics in two and four dimensions 
[1,2,3]. Ooguri and Vafa have shown that SdG is actually an effective theory of N=2 Strings, and that it has a most natural interpretation in the 
context of stringy physics [4]. The N=2 Heterotic String Theory (SdG coupled 
with Self-dual Yang-Mills theory (SdYM)) can be seen as the ``master 
system'' in which one is able to understand the deep interelation between the 
integrability in four dimensions (coming from the Penrose construction [5]) 
and the integrability arising in the two dimensional models [4]. It is hoped 
that the {\it quantum geometry} (geometry associated to the relevant 
$N=2$ superconformal field theory) coming form N=2 Heterotic Superstrings will 
generalize the Ricci-flat K\"ahler geometry involved in
 SdG, since SdG is conjectured to be related in a deep way to the quantum geometry trough the N=2 supersymmetric non-linear sigma models on Calabi-Yau manifolds [4] (for a recent review about quantum geometry see [6]). This quantum geometry might be usefu

l in order to understand (at the 
algebraic geometry level) the relation of the two twistor constructions 
associated with SdG and SdYM. This is because many of the features of the {\it classical} geometry are valid at {quantum} level.

All string theories are Conformal Field Theories in two dimensions 
(CFT2's). Almost all the `magic' properties of string theory  have their origin in 
this CFT2. These symmetries are expressed through the Virasoro,  
Affine Lie and {\it W}$_{\infty}$-algebras and they can be very 
useful in inducing (up to dimensional obstructions) nice 
properties to the self-dual structures in four dimensions.  One of these 
properties is the 
existence of conserved quantities in a physical theory. In  a CFT2 (due 
to the infinite dimensional symmetries) always arise infinite 
hierarchies of conserved quantities.

On the other hand in SdG the history has been a bit different. 
The first 
notion of an infinite hierarchy of conserved quantities has been given in a paper by 
Boyer and Pleba\'nski [7]. Using the first and the second heavenly 
equations [8] it is shown that SdG admits an infinite hierarchy of conservation 
laws. These quantities are defined in terms of the 
first and second key holomorphic functions $\Omega$ and $\Theta$. Later, some global aspects associated to the above  
construction have been studied [9]. For this the maximal isotropic submanifolds formalism is employed. This construction is formal, thus avoiding the 
use of 
the infinitesimal deformation of the twistor space [5]. The symmetries 
play an 
important role here, showing that the underlying symmetry group is the 
area preserving diffeomorphisms of a twistor surface (totally geodesic 
null surface). In fact, this approach shows the existence of a 
correspondence between the formal holomorphic bundles over the Riemann 
sphere ${\bf CP}^1$ and the group of area preserving diffeomorphisms. After 
this, Takasaki, in a series of papers [10], shows the 
existence of a hyper-K\"ahler infinite hierarchy. He found that it was  
possible to construct inequivalent metrics in SdG by using the area 
preserving diffeomorphisms group.

Later, Strachan has shown that the existence of the infinite hierarchy is  
related with an infinite family of twistor surfaces [11]. To be more 
precise, 
he found a one to one correspondence between a family of twistor 
surfaces and the conserved charges. Similarly to [7,9], Strachan  
starts also from the heavenly equation to make his construction.

Another approach to this setting was given recently by Husain [12]. He 
has used 
strongly his own result concerning the use of the equivalence of SdG and 
the two-dimensional chiral model with the gauge group defined by the group of 
area  
preserving diffeomorphism of an ``internal'' two-surface ${\cal N}^2$ [13]. Husain's 
construction of the infinite hierarchy involves the use of 
induced properties of the two-dimensional chiral model to SdG. In the 
present paper we continue  this philosophy and explore how some other 
features of a more general two-dimensional field theory (i.e., a CFT2) can be 
carry over to self-dual structures in four-dimensions.

In Sect. 2 we briefly review Husain's construction of the infinite 
hierarchy in SdG [12,14]. After this in Sect. 3 we show how this hierarchy has associated a hidden infinite Hopf algebra structure. Finally, in Sect. 4 our final remarks are given.

\vskip 1truecm

\chapter{\bf Infinite  Hierarchy of Conserved Currents in 
Self-dual Gravity}
\section{Conservation Laws From Husain's Chiral Model}

In this section we shall briefly review the necessary arguments in order 
to display the Hopf algebra structure of  sdiff$({\cal N}^2)$ in the next section. 
In Ref. [12,13], starting from the Ashtekar-Jacobson-Smolin (AJS) formulation for SdG [15], Husain found a set of equations for four vector fields ${\cal 
U}$, ${\cal V}$, ${\cal X}$ and ${\cal T}$. These vector fields arise 
after a light-cone variable decomposition of $V_0$ and a triad of vector fields 
$V_i$, $i=1,2,3$ on a three manifold $\Sigma^3$. $V_0$ is a vector field used in the $3+1$ splitting. Here $\Sigma^3$ comes 
from the global splitting of the space-time manifold ${\cal M}$ into 
$\Sigma^3 \times \Re$. The four-manifold ${\cal M}$ admits local 
coordinates $\{x^0,x^1,x^2,x^3\}$. In fact for the most relevant part of 
Husain's construction the three manifold $\Sigma^3$ can be identified 
locally with $\Re^3$. Choosing suitable expressions for the vector 
fields ${\cal U}$, ${\cal V}$, ${\cal X}$ and ${\cal T}$, Husain proved 
that the AJS equations led to the two-dimensional chiral model (with  
the two-dimensional `space-time' ($\Re^2$) coordinates $\{x,t\}$ and with gauge group being 
the group of area-preserving diffeomorphisms of a two-dimensional 
`internal space' ${\cal N}^2$\foot{As we make only 
local considerations we assume the space ${\cal N}^2$ to be a 
two-dimensional simply 
connected symplectic manifold with local coordinates $p$ and $q$. This space has a natural local 
symplectic structure given by the local area form $\omega = dp\wedge dq$. 
The 
group SDiff$({\cal N}^2)$ is precisely the group of diffeomorphisms on 
${\cal N}^2$ preserving the symplectic structure $\omega$,  i.e. for all $g\in 
{\rm SDiff}({\cal N}^2)$, $g^*(\omega) = \omega$.} with local coordinates $\{p,q\}$). This chiral model is 
given by the equations 

$$ f_{01}=\partial_0 a_1 - \partial_1a_0 + \{a_0,a_1\}_P= 0, 
\eqno(2.1a)$$

$$ \partial_0 a_0 + \partial_1 a_1  = \partial_i a_i 
= 0, \eqno(2.1b)$$
where $i=0,1$ (or $x,t$), $\{,\}_P$ means the Poisson bracket in $p$ and 
$q$ and $a_i = a_i(x,t,p,q)$ are analytic functions on ${\cal Y} = \Re^2 \times {\cal N}^2$ which satisfy the above 
equations. In the Husain's formulation the functions $a_i$'s are close 
related to the hamiltonian vector fields ${\cal U}^a$ and ${\cal V}^a$ 
which are given by 

$$ {\cal U}^a = \bigg({\partial \over \partial t}\bigg) + \omega^{ba} 
\partial_b H_0, \eqno(2.2a)$$

$$ {\cal V}^a = \bigg({\partial \over \partial x}\bigg) + \omega^{ba} 
\partial_b H_1, \eqno(2.2b)$$
where $H_0,H_1 \in C^{\infty}({\cal Y})$ are the hamiltonian functions which differ from 
$a_0$ and $a_1$ by the arbitrary functions $G$ and $F$ respectively [12]

$$  a_0 = H_0 + G, \ \ \ \ \ \ \ \ \ \ \  a_1 = H_1 - F. \eqno(2.3) $$
$\omega^{ab} = \big( {\partial \over \partial p} \big)^{[a} \otimes 
\big({\partial  \over \partial q} \big) ^{b]}$ whose inverse is precisely 
the symplectic local form $\omega_{ab} = dp_a \wedge dq_b$ on ${\cal N}^2$ and satisfying the relation $\omega^{ab} \omega_{bc} = \delta^a_c$. In all that $a,b= p,q$. All these considerations are of course local\foot{ Globally the symplectic form is defin
ed by $\omega : T{\cal N}^2 \to T^*{\cal N}^2$ and inverse $\omega^{-1}:T^*{\cal N}^2 \to T{\cal N}^2$. While the hamiltonian vector fields are ${\cal U}_{H_i} = \omega^{-1}(dH_i)$ satisfying the algebra $[{\cal U}_{H_i},{\cal U}_{H_j}] = {\cal U}_{\{H_i,
H_j\}}$ where $\{, \}$ stands for the Poisson bracket. Locally it can be written as $\{H_i,H_i\} = \omega^{-1}(dH_i,dH_j) = \omega^{ab} \partial_a H_i \partial_b H_j$.}.

Now we define a two-dimensional vector 
field-valued 1-form with `space-time' components precisely the vector fields ${\cal U}$ and ${\cal V}$, that is

$$ {\cal A}_i = ({\cal U} dt + {\cal V} dx)_i. \eqno(2.4) $$
where the notation means ${\cal A}_0 = {\cal A}_t= {\cal U}$ and ${\cal A}_1 = {\cal A}_x = {\cal V}$.

Thus the 1-form   $ {\cal A} = {\cal A}_i dx^i$ can be interpreted as a sdiff$({\cal N}^2)$-valued connection 1-form on $\Re^2$, {\it i.e.} ${\cal A} \in C^{\infty} \big( T^*\Re^2 \otimes {\rm sdiff}({\cal N}^2) \big)$. This connection is of course {\it f
lat} and that condition implies that its curvature $\Omega$ vanishes
 
$$  \Omega = d{\cal A} + {\cal A} \wedge {\cal A} = 0. \eqno(2.5)$$
Locally this condition yields
$$ {\cal F}_{01}= \partial_0 {\cal A}_1 -  \partial_1 {\cal A}_0 + [{\cal A}_0,{\cal A}_1] = 0, \eqno(2.6a)$$
where $[,]$ stands for the Lie bracket.

According to the Ref. [12]  the vector fields ${\cal U}$ and ${\cal V}$ must satisfy also the relation

$$ {\partial {\cal U} \over \partial t} + {\partial {\cal V} \over \partial x} = 0$$
or 
$$ \partial_i {\cal A}_i = \partial_0{\cal A}_0 + \partial_1 {\cal A}_1 = 0. \eqno(2.6b)$$

Thus in terms of the connection ${\cal A}_i$, the chiral Eqs. $(2.1a,b)$ can be expressed as Eqs. $(2.6a,b)$ respectively. 

The first conserved current is taken to be 

$$ {\cal J}_i^{(1)}(x,t,p,q) := {\cal A}_i(x,t,p,q), \eqno(2.7)$$
which is immediately seen to be conserved using the equation of motion 
$(2.6b)$. Introducing the two-dimensional Levi-Civita symbol 
$\epsilon_{ij}$ ($\epsilon_{01} = - \epsilon_{10} = 1$), it is easy to see 
that the first conserved current is

$$ {\cal J}_i^{(1)} = \epsilon_i^{ \ j} \partial_j \eta^{(1)}, \eqno(2.8)$$
implying the existence of a vector field $\eta^{(1)}$. 

The second conserved current is defined by 

$$ {\cal J}_i^{(2)} := \big[{\cal A}_i, \eta^{(1)}\big]. \eqno(2.9)$$
This current will be 
conserved using the Eqs. $(2.6a,b)$. Similarly to the above 
equation the $n$-th conserved current can be defined to be

$$ {\cal J}_i^{(n)} := \big[{\cal A}_i, \eta^{(n-1)}\big], \eqno(2.10)$$
which will be also conserved using the above argument.

Using now mathematical induction after several steps one can prove  that the $(n+1)$-th current

$$ {\cal J}_i^{(n+1)} := \big[{\cal A}_i, \eta^{(n)}\big], \eqno(2.11)$$
is also conserved (for details see Ref. [12]).

In what follows we will use only the first conserved current. The consideration of higher order conserved currents remain to be addressed.

The conserved charges $Q^{(1)}(t)$ and $Q^{(2)}(t)$ can be 
defined as:

$$
  Q^{(1)}(t)= \int_{\Re \times {\cal N}^2} d^2 \bar{s } \ dx {\cal J}_0^{(1)}(t,x,\bar{s}),
\eqno(2.12a)
$$
where $\bar s$ are the local coordinates on ${\cal N}^2$ ({\it i.e.} $d^2 \bar s = dp dq$) and ${\cal J}_0^{(1)} = {\cal A}_0 = {\cal U}$, then

$$
  Q^{(1)}(t) =  \int_{\Re \times {\cal N}^2} d^2 \bar{s} \ dx \ {\cal 
U}(t,x,p,q)\eqno(2.12b) $$ 
and

$$
     Q^{(2)}(t)= \int_{\Re \times {\cal N}^2} d^2 \bar{s} \  dx {\cal J}_0^{(2)}(t,x,\bar{s}), \eqno(2.13a)
$$

$$
     Q^{(2)}(t) = \int_{\Re \times {\cal N}^2} d^2 \bar{s} \ dx [{\cal U}, \int ^x dx' 
{\cal U}(t,x',p,q)]. \eqno(2.13b)
$$

\section{Affine Lie Algebra Associated to the Lie algebra of sdiff$({\cal N}^2)$}

In Ref. [14] Husain found an affine Lie algebra (of the Kac-Moody type) associated with the Lie algebra of area preserving diffeomorphisms. Beginning from the sdiff$({\cal N}^2)$-chiral equations $(2.1a,b)$ Husain show the existence of an infinite dimensi
onal hidden non-local symmetry associated with the Poisson bracket. The conservation law

$$ \partial_i J^{(n)}_i(x,t,p,q) = 0 \eqno(2.14)$$
for the corresponding currents of Eq. (2.10)

$$  J^{(n)}_i(x,t,p,q) = \epsilon_{ij} \partial_j \Lambda^{(n+1)} (x,t,p,q) \eqno(2.15)$$
determines the existence of a scalar function $\Lambda^{(n+1)}$ on $ {\cal Y} = \Re^2 \times {\cal N}^2$. This function can be obtained from a hierarchy of such a functions by

$$ \Lambda^{(n+1)} = \int_{- \infty}^x dx' D_0 \Lambda^{(n)}(t,x') \eqno(2.16)$$
where $D_0$ is the zero component of the covariant derivative $D_i \Lambda := \partial_i \Lambda + \{ a_i, \Lambda \}_P.$

The currents (2.15) are conserved under the symmetry transformations

$$\delta_{\Lambda} a_i = D_i \Lambda. \eqno(2.17)$$

Now, in order to obtain the above mentioned Kac-Moody algebra Husain defines the generators associated to the transformation (2.17) by

$$T^{(n)} := \int dt dx (\delta_{\Lambda}^{(n)} a_i) {\delta \over \delta a_i}$$
or

$$ = \int dt dx (D_i \Lambda^{(n)}) {\delta \over \delta a_i}. \eqno(2.18)$$

We first wish to give some general remarks in order to fix the notation 
for further considerations. For this we will use the standard notation of 
general 2-index infinite algebras [16,17,18]. To be more precise let 
$\{ {\bf e_m(x)}\}$ be a generic basis of hamiltonian functions satisfying the algebra

$$\{ {\bf e_m(x)}, {\bf e_m'(x)} \} =  C^{\bf m''}_{\bf m m'} {\bf e_m''(x)}. 
\eqno(2.19)$$
Expanding now $\Lambda^{(n)}(x,t,p,q)$ in the above basis $\{ {\bf e_m(x)}\}$ we have

$$ \Lambda^{(n)}(x,t,p,q) = \sum_{\bf m}  {\bf e_m(x)} \Lambda^{(n)}_{\bf m}(x,t) \eqno(2.20)$$
where ${\bf m}, {\bf m'}$ and ${\bf m''}$ are constant 2-vectors, ({\it 
i.e.} ${\bf m} = (m_1,m_2)$, with $m_1,m_2 \in {\bf Z}$ and ${\bf x} = 
(p,q)$ is a 2-vector with $p,q$ the local coordinates on ${\cal N}^2$) 
and  $C^{\bf m''}_{\bf mm'}$ are the structure constants which depend on 
the topology of ${\cal N}^2$.

Let ${\cal L}_{\bf e_m(x)} \equiv  {\cal L}_{\bf m}$ be the associated 
hamiltonian vector fields which satisfy the Poisson algebra sdiff$({\cal 
N}^2)$

$$ [{\cal L}_{\bf m}, {\cal L}_{\bf m'}] = C^{\bf m''}_{\bf mm'} {\cal L}_{\bf m''}. \eqno(2.21)$$
Any general function ${\cal F}({\bf x})$ 
can be expressed as the linear combination of the basis of the vector fields

$$ {\cal F}({\bf x}) = \sum_{\bf m} f_{\bf m} {\cal L}_{\bf m}, \eqno(2.22)$$
where $f_{\bf m}$ are the expansion coefficients.

In terms of the basis $\{ {\bf e_m(x)} \}$ the generators $T^{(n)}$ can be expressed by

$$ T^{(n)}_{\bf m} = \int dt dx (D_i \Lambda^{(n)}_{\bf m}) {\delta \over \delta a_i}. \eqno(2.23)$$

Finally Husian found also an affine Lie algebra structure for these generators $T^{(n)}_{\bf m}$ to be

$$ [T^{(m)}_{\bf m}, T^{(n)}_{\bf m'}] = C_{\bf mm'}^{\bf m''} T^{(m+n)}_{\bf m''}. \eqno(2.24) $$

The most trivial case, $m=n=0$, corresponds precisely with the Poisson algebra sdiff$({\cal N}^2)$ (2.21)

$$ [T^{(0)}_{\bf m}, T^{(0)}_{\bf m'}] = C_{\bf mm'}^{\bf m''} T^{(0)}_{\bf m''}. \eqno(2.25) $$

These results are of course at the classical level. The quantization might be related to knot theory and integrable 2d field theory. Accoeding with Ref. [14] it is possible the this connection come from the Yang-Baxter equation.

\vskip 1truecm

\chapter{\bf A Hopf Algebra Structure in Self-dual Gravity}

In this section we make the construction of the Hopf algebra ${\cal 
H}$ associated to the affine Lie algebra (2.24), in particular, that associated to the Poisson algebra sdiff$({\cal N}^2)$.

We define new generators (or charges) ${\cal Q}^{(n)}_{\bf m}(t)$ from Eq. (2.23) by

$$ T^{(n)}_{\bf m} = \int dt \ {\cal Q}^{(n)}_{\bf m}(t)  \eqno(3.1)$$
where
$$  {\cal Q}^{(n)}_{\bf m}(t)= \int_{- \infty}^{\infty} dx \ j^{(n)}_{\bf m}(x,t) \eqno(3.2)$$
and 
$$ j^{(n)}_{\bf m}(x,t) = D_i \Lambda^{(n)}_{\bf m}(x,t) {\delta \over \delta a_i}. \eqno(3.3)$$

The defined charges satisfy of course the algebra

$$ [{\cal Q}^{(m)}_{\bf m}(t), {\cal Q}^{(n)}_{\bf m'}(t')] = \delta(t -t') C_{\bf mm'}^{\bf m''} {\cal Q}^{(m+n)}_{\bf m''}(t). \eqno(3.4) $$

Now we decompose the integral (3.2) into two integrals

$$  {\cal Q}^{(n)}_{\bf m}(t)= \int_{- \infty}^{0} dx \ j^{(n)}_{\bf m}(x,t) +  \int_{0}^{\infty} dx \ j^{(n)}_{\bf m}(x,t). \eqno(3.5)$$

\noindent
Thus we can write the last equation as
$$
 {\cal Q}^{(n)}_{\bf m}(t)= {\cal Q}^{(n)}_{\bf m \ +}(t) + {\cal Q}^{(n)}_{\bf m \ -}(t). \eqno(3.6) $$
where the signs $+(-)$ correspond to positive (negative) values of $x$.

In what follows we restrict ourselves to the case (2.25) ({\it i.e} $m=n=0$), but the generalization for the general case is easy to get.

\section{The Hopf Algebra Structure}

It is well known that the diffeomorphism group Diff$(M)$ of a compact 
manifold $M$ is not a Banach Lie Group but a Hilbert manifold 
[19,20]. Due to that ${\cal N}^2$ is a compact symplectic manifold with symplectic form $\omega = dp \wedge dq$ which defines a 
subgroup SDiff$({\cal N}^2)$ of Diff$({\cal N}^2)$. The associated Lie algebra to 
SDiff$({\cal N}^2)$ is sdiff$({\cal N}^2)$, the space of locally hamiltonian vector 
fields on ${\cal N}^2$. This algebra has the structure of a Frechet manifold and therefore the product sdiff$({\cal N}^2) \times$ sdiff$({\cal N}^2)$ is the product of Frechet manifolds which is also Frechet [20]. 

Consider the infinite dimensional Lie algebra sdiff$({\cal N}^2)$ over the 
field $\Re$ and a basis of this algebra to be the ${\cal Q}^{(0)}_{\bf m}(t)$. 
Introducing an 
infinite dimensional Universal Enveloping algebra ${\cal H}= {\bf 
U}\big({\rm sdiff}({\cal N}^2)\big)$, we can now define on ${\cal H}$ a 
structure of a Hopf algebra (for details see Ref. [21]). Thus, in what follows the tensorial product of Universal Enveloping 
algebras of sdiff$({\cal N}^2)$, ${\cal H}\otimes {\cal H}$, will be defined in 
the Frechet sense.

Following MacKay's [22] we define the co-product, $\Delta$, as follows:

$$\Delta : {\cal H} \to {\cal H}\otimes {\cal H}, \eqno(3.7)$$
that means,

$${\cal Q}^{(0)}_{\bf m} \mapsto \Delta({\cal Q}^{(0)}_{\bf m}) = {\cal 
Q}^{(0)}_{\bf m}\otimes 1 + 1\otimes {\cal Q}^{(0)}_{\bf m}. \eqno(3.8)$$
On the identity element, $1\in {\cal H}$, the co-product is defined by

$$\Delta(1) = 1\otimes 1. \eqno(3.9)$$
 
\noindent
This co-product $\Delta$ is an $\Re$-algebra homomorphism. The 
definition for the co-product, Eq. (3.8), is fulfilled also when we 
define 

$$ {\cal Q}^{(0)}_{\bf m}(t) = {\cal Q}^{(0)}_{{\bf m} \ 1+}(t) \otimes  {\cal 
Q}^{(0)}_{{\bf m} \ 2-}(t), \eqno(3.10)$$
where the signs $+$ and $-$ correspond to the decomposition given in Eq. 
(3.6), the numbers 1,2 mean the first and the second entries of ${\cal 
H}\otimes {\cal H}$, respectively.

Defining now the `twist' map $\tau: {\cal H}\otimes {\cal H} \to {\cal 
H}\otimes {\cal H}$ given by $\tau\big({\cal Q}^{(0)}_{\bf m} \otimes {\cal 
Q}^{(0)}_{\bf m'}\big) = {\cal Q}^{(0)}_{\bf m'} \otimes {\cal Q}^{(0)}_{\bf m}$, one 
can see that the co-product (3.8) is co-commutative. This holding 
because the relation $\tau \circ \Delta = \Delta$ is fulfilled.

With the the above definition one can show that the coproduct satisfies 
the co-associativity axiom

$$(id \otimes \Delta)\circ \Delta = (\Delta \otimes id)\circ \Delta. 
\eqno(3.11)$$
where $id$ is the identity map, {\it i.e.} $id: {\cal H} \to 
{\cal H}$, 
$id({\cal Q}^{(0)}_{\bf m}(t)) = {\cal Q}^{(0)}_{\bf m}(t)$. To prove this axiom one can  
decompose the charge ${\cal Q}^{(0)}_{\bf m}(t)$ into three parts just as it 
is mentioned in Ref. [22].

On the other hand the co-unit $\epsilon$ is also  an $\Re$-algebra 
homomorphism $\epsilon: {\cal H} \to \Re$, which one can define by

$$
\epsilon({\cal Q}^{(0)}_{\bf m}(t)) = 0, \ \ \ \  \epsilon(1) = 1.\eqno(3.12)
$$
where $0 \in \Re$.
With the above definitions for the co-product and co-unit one can easily 
prove that the co-unit axiom is fulfilled. 

$$(id \otimes \epsilon)\circ \Delta = (\epsilon \otimes id)\circ \Delta. 
\eqno(3.13)$$

The antipode is an $\Re$-algebra antihomomorphism $S: {\cal H} \to 
{\cal H}$. In our case we define the antipode as

$$
S({\cal Q}^{(0)}_{\bf m}(t)) = - {\cal Q}^{(0)}_{\bf m}(-t). \eqno(3.14)
$$
It is an easy matter to see that the definitions (3.8) and (3.14) satisfy the axiom 
of the antipode:

$$m\circ(S\otimes id)\circ \Delta = m\circ (id \otimes S)\circ \Delta, 
\eqno(3.15)$$
where $m$ is the operation product in ${\cal H}$ and is defined as a 
homomorphism $m: {\cal H}   \otimes {\cal H} \to {\cal H}$, $m \big( 
{\cal Q}^{(0)}_{\bf m}\otimes {\cal Q}^{(0)}_{\bf m'} \big):=  [{\cal Q}^{(0)}_{\bf m}, {\cal Q}^{(0)}_{\bf m'} ] = C^{\bf m''}_{\bf m m'}{\cal Q}^{(0)}_{\bf m''}$, where 
${\cal Q}^{(1){\bf m}}, {\cal Q}^{(1){\bf m'}} \in {\cal H}$ 
 
From the point of view of current algebra the above 
definition for the antipode map, (3.14), corresponds to a usual, classical,  {\bf PT} transformation

$$S\big(j^{(0)}_{\bf m}(x,t)\big) = - j^{(0)}_{\bf m}(-x,-t). 
\eqno(3.16)$$
which coincides with McKay's result [22] whenever 

$$ S\big( \Lambda^{(n)}_{\bf m}(x,t) \big) = \Lambda^{(n)}_{\bf m}(-x,-t). \eqno(3.16)$$

\vskip 1truecm

\chapter{ \bf Final Remarks}

In the present paper we have used the infinite hierarchy of conserved 
quantities for SdG just as it has been given by Husain [12,14]. Then using the conserved charges we display how they possess a Hopf algebra structure. Many interesting implications might be derived from our results. First of all one could generalize every
thing presented here by considering the Moyal algebra. This is the unique deformation (with deformation parameter to be $k$) of the Poisson algebra considered here sdiff$({\cal N}^2)$. The Moyal algebra is 

$$ [ {\cal L}_{\bf m}, {\cal L}_{\bf m'}] = {1 \over k} Sin \bigg( k {\bf m \times m'} \bigg) {\cal L}_{\bf m + m'}. \eqno(4.1)$$
One would like to find the Hopf algebra associated to the Moyal deformation of sdiff$({\cal N}^2)$. In fact there exist a further $q$-deformation of the Moyal algebra. It was considered in Refs. [23,24,25] a $q$-deformation of the Moyal algebra
$$[{\cal L}_{\bf m}, {\cal L}_{\bf m'}]_{q^{\bf m \times m'}} = \bigg( p^{\bf m\times m'} - p^{- {\bf m \times m'}}\bigg) {\cal L}_{\bf m + m'}. \eqno(4.2)$$
This structure of quantum algebra of the Moyal deformation of sdiff$({\cal N}^2)$  might be important in order to give a bit of more consistency to our results [26].

Another way to address a quantum algebra structure is trough the existence of a QUE (Quantized Universal Enveloping)-algebra associated with both sdiff$({\cal N}^2)$ and its Moyal deformation might be achieved using the methods of massive 2d 
quantum field theory given by Le Clair and Smirnov [27,28]. Then by considering these methods one will have a no(co)commutative (co)product which give us directly a $q$-deformed version of sdiff$({\cal N}^2)$ and its Moyal deformation.

One more possible things to be considered in this context concerning to conserved currents. Strachan found an infinite hierarchy of 
symmetries associated to SdG equations [29]. A different approach has been given also very recently by Husain [14]. It would be interesting  to make the connection between both approaches which seem to be equivalent. 

Furthermore, since the Husain's model has been solved (for finite subgroups 
of sdiff$({\cal N}^2)$) in terms of harmonic maps [30], one could ask about the 
possibility to address the whole problem concerning conserved laws, Hopf 
algebras, its deformations and harmonic maps.

Finally, given the possible relations between self-dual gravity and knot theory by using the Yang-Baxter equation one could to use the technique [31] in order to get such a connections. This technique has ben succeful to find solutions in self-dual gravit
y and to obtain a like-WZW action which might be connected to 2d integrable field theory. We would like to address this considerations in a forthcoming paper.

\vskip 1truecm

\centerline{\Lina Acknowledgments}
We would like to thank J.D. Finley III and M.  Przanowski for helpful  discussions  and suggestions and also to M.B. Halpern for pointing us out the correct use of the functional names for the different algebras. This work was supported in part by CONACyT
 and SNI.

\vskip 2truecm
\centerline{\Hugo References}

\item{1.} M.F. Atiyah, Commun. Math. Phys. {\bf 93} (1984) 437.

\item{2.} Q-Han Park, Phys. Lett. {\bf B257} (1991) 105; {\bf 
B238} (1990) 287; Int. J. Mod. Phys. {\bf A7} (1992) 1415.

\item{3.} I. Bakas, `Area Preserving Diffeomorphism and Higher Spin 
Fields in Two Dimensions' Proceedings of the Trieste Conference on 
Supermembranes and Physics in 2+1 Dimensions, Eds. M. Duff, C. Pope and 
E. Sezgin, World Scientific (1990) 352-362; Commun. Math. Phys. {\bf 
134} (1990) 487; Int. J. Mod. Phys. {\bf A6} (1991) 2071; ``Self-duality, Integrable Systems, W-algebras and 
All That'', in: {\it Nonlinear Fields: Classical, Random, Semiclassical}, 
eds. P. Garbaczewski and Z. Popowicz (World Scientific Publishings Co. 
Pte. Ltd, 1991) pp. 2-35.

\item{4.} H. Ooguri and C. Vafa, Nucl. Phys. 
{\bf B361} (1991) 469; Mod. Phys. Lett. {\bf A5} (1990) 1389; Nucl. Phys. {\bf B367} (1991) 83; Nucl. Phys. {\bf B451} (1995) 121.

\item{5.} R. Penrose, Gen. Rel. Grav. {\bf 7} (1976) 31.

\item{6.} B.R. Green, ``Lectures on Quantum Geometry'', Nucl. Phys. {\bf B} (Proc. Suppl.) {\bf 41} (1995) 92.

\item{7.}C.P. Boyer and J.F. Pleba\'nski, J. Math. Phys. {\bf 18} (1977) 1022.

\item{8.} J.F. Pleba\'nski, J. Math. Phys.  {\bf 16} (1975) 2395.

\item{9.} C.P. Boyer and J.F. Pleba\'nski, J. Math. Phys. {\bf 26} (1985) 229; C.P. Boyer, in {\it Non-linear Phenomena}, Ed by K.B. Wolf, 
Lecture Notes in Physics, {\bf 189}, Springer-Verlag (1983).

\item{10.} K. Takasaki, J. Math. Phys. {\bf 30} (1989) 1515; J. Math. 
Phys. {\bf 31} (1990) 1877; Commun. Math. Phys. {\bf 127} (1990) 255;  `Area Preserving Diffeomorphisms and Non-linear 
Integrable Systems' in {\it Topological and Geometrical Methods in Field Theory} Eds. J. Mickelsson and O. Pekonen, World Scientific, Singapore (1992); Phys. Lett. {\bf B285} (1992) 187;{\it W Algebra, Twistor and Nonlinear Integrable 
Systems}, Preprint KUCP-0049/92, June 1992.

\item{11.} I.A.B.  Strachan, Class. Quantum Grav. {\bf 10} (1993) 1417.

\item{12.} V. Husain, Class. Quantum Grav. {\bf 11} (1994) 927.

\item{13.} V. Husain, Phys. Rev. Lett. {\bf 72} (1994) 800.

\item{14.} V. Husain,  J. Math. Phys. {\bf 36} (1995) 6897.

\item{15.} A. Ashtekar, T. Jacobson and L. Smolin, Commun. Math. Phys. 
{\bf 115} (1989) 631.

\item{16.} I.M. Gelfand and I. Ya Dorfman, Funk. Anal. Pril. {\bf 13} (1980) 248; {\bf 14} (1990) 248; {\bf 15} (1981) 173; {\bf 16} (1982) 241.

\item{17.} E.G. Floratos, J. Iliopoulos and G. Tiktopoulos, Phys. Lett. {\bf B217} (1989) 285.

\item{18.} V.I. Arnold, {\it Mathematical Methods of Classical Mechanics}, 
Springer-Verlag, (1978). 

\item{19.} J. Milnor, `Remarks on Infinite-dimensional Lie Groups' in 
Les Houches, 1983; {\it Relativity Groups and Topology II}, eds. B.S. De Witt 
and R. Stora. Elsevier Science Publishers B.V., (1984).

\item{20.} R. Schmid, {\it Infinite Dimensional Hamiltonian Systems}, 
Bibliopolis, (1987).

\item{21.} L.A. Takhtajan, Nankai lecture notes (1989), in: {\it 
Introduction to Quantum Groups and Integrable Massive Models of Field 
Theory} eds. Mo-Lin Ge and Bao-Heng Zhao (World Scientific, Singapore, 
1990), p. 69.

\item{22.} N.J. MacKay, Phys. Lett. {\bf B 281} (1992) 90.

\item{23.} C. Zachos, `` Paradigms of Quantum Algebras'', Preprint 
ANL-HEP-PR-90-61, Updated, January 1992; Preprint ANL-HEP-CP-89-55.

\item{24.} D.B. Fairlie, `Polynomial Algebras with $q$-Heisenberg 
Operators', in the {\it Proceedings of the Argonne Workshop 
on Quantum Groups}, Eds. T. Curtright, D.B. Fairlie and C. Zachos, World 
Scientific, Singapore 1991.

\item{25.} P. Fletcher, `The Moyal Bracket', in the {\it Proceedings of the 
Argonne Workshop 
on Quantum Groups}, Eds. T. Curtright, D.B. Fairlie and C. Zachos, World 
Scientific, Singapore 1991.

\item{26.} J.F. Pleba\'nski and H. Garc\'{\i}a-Compe\'an, Int. J. Mod. Phys. {\bf A10} (1995) 3371.

\item{27.} A. LeClair and F.A. Smirnov, Int. J. Mod. Phys. {\bf A7} (1992) 2997 .
\item{28.} A. LeClair, ``Infinite Quantum Group Symmetry in 2D Quantum 
Field Theory'', in the Proceedings of the XXth International Conference on Differential Geometric Methods in Theoretical Physics, New York City.

\item{29.} I.A.B. Strachan, J. Math. Phys. {\bf 36} (1995) 3566.

\item{30.} H. Garc\'{\i}a-Compe\'an and T. Matos, Phys. Rev. D {\bf 52} (1995) 4425.

\item{31.} J.F. Pleba\'nski, M. Przanowski and H. Garc\'{\i}a-Compe\'an, `From Principal Chiral Model to Self-dual Gravity', unpublished, hep-th/9509092; `Further Remarks on the Chiral Model Approach to Self-dual Gravity', unpublished, hep-th/9512013.

\endpage

\end